\documentstyle [times,epsf,11pt]{article}

\def\pt  {$p_{T}$}
\def\xf  {$x_{F}$}

\def\eplus {$e^{+}$}
\def\emin {$e^{-}$}

\def\piplus {$\pi^{+}$}
\def\pimin {$\pi^{-}$}
\def\pizero {$\pi^{0}$}

\def\kplus {K$^{+}$}
\def\kmin {K$^{-}$}
\def\kzero {K$^{0}$}

\def\sigmaplus {$\Sigma^{+}$}
\def\sigmamin {$\Sigma^{-}$}
\def\sigmazero {$\Sigma^{0}$}

\def\lambdazero {$\Lambda^{0}$}
\def\antilambda {$\overline{\Lambda}^{0}$}

\def\xipp {$\Xi^{0}(1530)$}
\def\xiss {$\Xi^{0}(1690)$}
\def\xizero {$\Xi^{0}$}
\def\ximin {$\Xi^{-}$}

\def\omegamin {$\Omega^{-}$}

\def\xicharmplus {$\Xi_{c}^{+}$}
\def\xicharmzero {$\Xi_{c}^{0}$}

\def\omegacharmzero {$\Omega_{c}^{0}$}

\def\lambdacharmplus {$\Lambda_{c}^{+}$}

\def\kevc1{\ifmmode\mathrm{\ keV/{\mit c}}
          \else$\mathrm{\ keV/{\mit c}}$\fi}
\def\Mevc1{\ifmmode\mathrm{\ MeV/{\mit c}}
          \else$\mathrm{\ MeV/{\mit c}}$\fi}
\def\gevc1{\ifmmode\mathrm{ GeV/{\mit c}}
          \else$\mathrm{ GeV/{\mit c}}$\fi}
\def\kevc2{\ifmmode\mathrm{\ keV/{\mit c}^2}
          \else$\mathrm{\ keV/{\mit c}^2}$\fi}
\def\Mevc2{\ifmmode\mathrm{\ MeV/{\mit c}^2}
          \else$\mathrm{\ MeV/{\mit c}^2}$\fi}
\def\Gevc2{\ifmmode\mathrm{\ GeV/{\mit c}^2}
          \else$\mathrm{\ GeV/{\mit c}^2}$\fi}

\begin{document}
\Large
\vspace{0.8cm}
\normalsize
\noindent
\begin{flushright}
  CERN PPE 97-134\\
  MPI H-V38-1997\\
  October\ 9, 1997
\end{flushright}
\vspace{0.8cm}
\Large

\begin{center}           
             {First observation of the \ximin\ \piplus\ decay mode
             of the \xizero(1690) hyperon }       \\
\large
\vspace{0.8cm}
            {The WA89 Collaboration} \\
\vspace{0.4cm}
\end{center}
\normalsize

\noindent
\sloppy
  M.I.~Adamovich$^8$,
  Yu.A.~Alexandrov$^8$,
  D.~Barberis$^3$,
  M.~Beck$^5$,
  C.~B\'erat$^4$,
  W.~Beusch$^2$,
  M.~Boss$^6$,
  S.~Brons$^{5,a}$,
  W.~Br\"uckner$^5$,
  M.~Bu\'enerd$^4$,
  C.~Busch$^6$,
  C.~B\"uscher$^5$,
  F.~Charignon$^4$,
  J.~Chauvin$^4$,
  E.A.~Chudakov$^{6,b}$,
  U.~Dersch$^5$,
  F.~Dropmann$^5$,
  J.~Engelfried$^{6,c}$,
  F.~Faller$^{6,d}$,
  A.~Fournier$^4$,
  S.G.~Gerassimov$^{5,8}$,
  M.~Godbersen$^{5}$,
  P.~Grafstr\"om$^2$,
  Th.~Haller$^5$,
  M.~Heidrich$^5$,
  E.~Hubbard$^{5}$,
  R.B.~Hurst$^3$,
  K.~K\"onigsmann$^{5,e}$,
  I.~Konorov$^{5,8}$,
  N.~Keller$^6$,
  K.~Martens$^{6,f}$,
  Ph.~Martin$^4$,
  S.~Masciocchi$^{5}$,
  R.~Michaels$^{5,g}$,
  U.~M\"uller$^7$,
  H.~Neeb$^{5}$,
  D.~Newbold$^1$,
  C.~Newsom$^{h}$,
  S.~Paul$^5$,
  J.~Pochodzalla$^5$,
  I.~Potashnikova$^5$,
  B.~Povh$^5$,
  Z.~Ren$^5$,
  M.~Rey-Campagnolle$^4$,
  G.~Rosner$^7$,
  L.~Rossi$^3$,
  H.~Rudolph$^{7}$,
  C.~Scheel$^i$,
  L.~Schmitt$^7$,
  H.-W.~Siebert$^6$,
  A.~Simon$^{6,e}$,
  V.J.~Smith$^{1,j}$,
  O.~Thilmann$^6$,
  A.~Trombini$^{5}$,
  E.~Vesin$^4$,
  B.~Volkemer$^{7}$,
  K.~Vorwalter$^5$,
  Th.~Walcher$^7$,
  G.~W\"alder$^{6}$,
  R.~Werding$^{5}$,
  E.~Wittmann$^5$, and
  M.V.~Zavertyaev$^{8}$

\vspace{0.2cm}
\begin{flushleft}
\noindent
    $^1${\sl University of Bristol, Bristol BS8 1TL, United Kingdom.}\\
    $^2${\sl CERN, CH-1211 Gen\`eve 23, Switzerland.}\\
    $^3${\sl Dipartimento di Fisica and I.N.F.N., Sezione di Genova, 
        I-16146 Genova, Italy.}\\
    $^4${\sl Institut des Sciences Nucl\'{e}aires, Universit\'{e} de Grenoble, 
        F-38026 Grenoble Cedex, France.}\\
    $^5${\sl Max-Planck-Institut f\"ur Kernphysik Heidelberg,
        D-69029 Heidelberg, Germany.}\\
    $^6${\sl Physikalisches Institut, Universit\"at Heidelberg,
        D-69120 Heidelberg, Germany.$^k$}\\
    $^7${\sl Institut f\"ur Kernphysik, Universit\"at Mainz,
        D-55099 Mainz, Germany.$^k$}\\
    $^8${\sl Moscow Lebedev Physics Institute,
        RU-117924 Moscow, Russia.}\\
\end{flushleft}
\vspace{0.2cm}

\normalsize
\centerline{\underline{\large{Abstract}}}
\vspace{0.5cm}

We report the first observation of the \ximin \piplus\ decay 
mode of the \xiss, confirming the existence of this resonance.
The \xiss\ were produced by $\Sigma^-$ of 345~\gevc1 mean momentum
in copper and carbon targets. The mass and width are close to those 
observed earlier for the $\Xi^-(1690)$ in the $\Lambda K^-$ decay channel.
The product of inclusive production cross section and branching ratio
is given relative to that of the $\Xi^0(1530)$.

\vspace{0.5cm}

\centerline{(Submitted to Zeitschrift f\"ur Physik C)}
\vspace{2.0cm}
\newpage

\setlength{\oddsidemargin}   {0.0in}
\setlength{\evensidemargin}  {0.0in}
\setlength{\textwidth}       {6.5in}
\setlength{\textheight}      {9.0in}
\setlength{\topmargin}      {-0.3in}
\setlength{\headheight}      {0.3in}
\setlength{\headsep}         {0.3in}
\setlength{\footskip}        {0.6in}
\setlength{\footheight}      {0.3in}
\vspace{4in}
\hrule width 2truein
\vspace{0.15cm}
\small
\noindent
    $^a${\sl Now at TRIUMF, Vancouver, B.C., Canada V6T 2A3.}\\
    $^b${\sl On leave of absence from Moscow State University,
             119889 Moscow, Russia.}\\
    $^c${\sl Now at FNAL, PO Box 500, Batavia,
          IL 60510, USA.} \\
    $^d${\sl Now at Fraunhofer Institut f\"ur Solarenergiesysteme,
               D-79100 Freiburg, Germany.}\\
    $^e${\sl Now at Fakult\"at f\"ur Physik, Universit\"at Freiburg, Germany.}\\
    $^f${\sl Now at Institute for Cosmic Ray Research, University of
             Tokyo,Japan.} \\
    $^g${\sl Now at Thomas Jefferson Lab, Newport News, VA 23606, USA.} \\
    $^h${\sl University of Iowa, Iowa City, IA 52242, USA.}\\
    $^i${\sl NIKHEF,1009 DB Amsterdam, The Netherlands.}\\
    $^j${\sl supported by the UK PPARC.}\\
    $^k${\sl supported by the Bundesministerium f\"ur Bildung, Wissenschaft,
          Forschung und Technologie, Germany, under contract numbers 
          05~5HD15I, 06~HD524I and 06~MZ5265.}\\
\normalsize

\clearpage

\pagestyle{plain}

\setlength{\oddsidemargin}   {0.0in}
\setlength{\evensidemargin}  {0.0in}
\setlength{\textwidth}       {6.5in}
\setlength{\textheight}      {9.0in}
\setlength{\topmargin}      {-0.3in}
\setlength{\headheight}      {0.3in}
\setlength{\headsep}         {0.3in}
\setlength{\footskip}        {0.6in}
\setlength{\footheight}      {0.3in}



\def\eplus {$e^{+}$}
\def\emin {$e^{-}$}

\def\piplus {$\pi^{+}$}
\def\pimin {$\pi^{-}$}
\def\pizero {$\pi^{0}$}

\def\kplus {K$^{+}$}
\def\kmin {K$^{-}$}
\def\kzero {$K^{0}$}

\def\sigmaplus {$\Sigma^{+}$}
\def\sigmamin {$\Sigma^{-}$}
\def\antisigmaplus {$\overline{\Sigma}^{-}$}
\def\sigmazero {$\Sigma^{0}$}

\def\lambdazero {$\Lambda^{0}$}
\def\antilambda {$\overline{\Lambda}^{0}$}

\def\xizero {$\Xi^{0}$}
\def\ximin {$\Xi^{-}$}
\def\antiximin {$\overline{\Xi}^{+}$}

\def\omegamin {$\Omega^{-}$}

\def\xicharmplus {$\Xi_{c}^{+}$}
\def\xicharmzero {$\Xi_{c}^{0}$}

\def\omegacharmzero {$\Omega_{c}^{0}$}

\def\lambdacharmplus {$\Lambda_{c}^{+}$}

\def\pt  {$p_{T}$}
\def\xf  {$x_{F}$}



\setcounter{page}{1}
\section{Introduction}
\label{sec_1}

More than three decades after the first observations of excited
states of hyperons, the excited states of $\Xi^-$ and $\Omega^-$
are still largely unexplored. Of the $\Xi^*$ states, only the $\Xi(1530)$
rates four stars in the PDG ranking, while four other states, 
the $\Xi(1690)$, $\Xi(1820)$, $\Xi(1950)$ and $\Xi(2030)$
rate three stars \cite{pdg}. 

First experimental evidence for the $\Xi(1690)$ came from a bubble chamber
experiment using a $K^-$ beam of 4.2~\gevc1. A strong threshold enhancement
was observed in the $\Sigma \overline{K}$ mass spectra, with weaker
evidence from the $\Lambda \overline{K}$ spectra \cite{DIO78}.
The first direct observation of the $\Xi(1690)$ as a resonance
resulted from a hyperon beam experiment at CERN. 
Here, a peak at 1690 \Mevc2 in $\Lambda K^-$ pairs produced diffractively
in a $\Xi^-$ beam of 116~\gevc1 was observed \cite{BIA87I}.
In an earlier run of that experiment, a corresponding signal
at around 1700 \Mevc2 with poorer mass resolution was seen \cite{BIA81}.

In the framework of the nonrelativistic quark potential model,
a $\Xi (1/2^+ )$ state was predicted with a mass around 1690 \Mevc2,
with dominating $\Xi \pi$ decay \cite{chao81}.
A relativistic version of this model, however, pushed the first excited 
$\Xi (1/2^+ )$ to about 1800 \Mevc2, and left no state to be identified with 
the observed $\Xi (1690)$ \cite{caps86}.
Also within a more recently developed chiral boson exchange 
interaction model the 
$\Xi (1/2^+ )$ state is expected at an energy far above 1690 \Mevc2 and 
close to 1800 \Mevc2 \cite{GLO96}.

In this paper, we report on a search for the $\Xi(1690)$ resonance 
in the $\Xi^- \pi^+$ channel. While in previous studies 
no statistically significant resonance signal around 1690 \Mevc2 
was observed in this decay mode \cite{BIA87I}, we find a clear 
resonant signal at a mass of 1686 \Mevc2. 

\section{The experimental apparatus}

The hyperon beam and the experimental setup were described in
detail elsewhere \cite{beam,expt}. Here we give only a brief summary
of the equipment important for this particular measurement.

The hyperon beamline selected \sigmamin\ hyperons with a mean momentum of 
345~\gevc1 and a momentum spread of $\sigma (p)/p=9\%$. 
Although the actual $\pi^-$ to $\Sigma^-$ ratio of the beam was
about 2.3, high-momentum pions were strongly suppressed on the trigger level by
a set of transition radiation detectors \cite{TRD} resulting in a remaining pion
contamination of about 12\%.
In addition the beam contained small admixtures of $K^-$ and $\Xi^-$ \cite{expt}. 
$\Sigma^-$ decays upstream of the experimental target provided a background of
neutrons and $\pi^-$ at lower energy, which could be rejected
by requiring that the beam track measured upstream of the target
intercepted the interaction vertex and fulfilled the position/angle
correlations given by the beam optics.
The trajectories of incoming and outgoing particles were measured
in silicon microstrip detectors upstream and downstream of the target.
The experimental target itself consisted of
one copper slab with a thickness of 0.025 $\lambda_I$ in beam direction,
followed by three carbon (diamond powder) slabs of 0.008 $\lambda_I$ each.

The momenta of the decay particles were measured in a magnetic spectrometer 
equipped with MWPCs and drift chambers. The spectrometer magnet was placed 
with its center 13.6 m downstream of the target to allow hyperons and $K^0_S$ 
emerging from the target to decay in front of it.

Charged particles could be identified using a ring imaging Cherenkov 
(RICH) detector \cite{BEU92}, which intercepted particles
with momenta above about 12~\gevc1.
In the analysis described below, particle identification was used for 
cross-check purposes only.

\section{Event selection}

The event selection for the decay chain 
$\Xi^* \rightarrow $ \ximin\ \piplus\ , 
\ximin\ $\rightarrow $ \lambdazero\ \pimin\ ,
\lambdazero\ $\rightarrow p$ \pimin\ was done as follows:

Combinations of positive and negative particles were
accepted as \lambdazero\ candidates 
if the distance of the two tracks at 
the decay point was smaller than 0.5 cm
and if their reconstructed $p\pi^-$ mass was within $\pm 3\sigma_1 $
of the \lambdazero\ mass.
Here $\sigma_1$ is the uncertainty of the 
mass determination based on the track properties of the individual events. 
Typically, $\sigma_1$ is about 1.6 \Mevc2.

\ximin\ candidates were accepted if their trajectory measured 
in the vertex detector downstream of the target agreed within 
errors with the \ximin\ momentum direction and the \ximin\ 
decay vertex which were constructed from the 
\lambdazero\ and $\pi^-$ daughther particles.
Furthermore, the reconstructed $\Lambda \pi^-$ mass
had to be within $\pm 3\sigma_2 $ of the \ximin\ mass where
$\sigma_2$ is typically 
2.7 \Mevc2 .

The \ximin\ production vertex had to contain at least one more charged
particle track besides the \ximin\ track.
The reconstructed vertex position had to be inside
a target block with an error margin of 3$\sigma$ in each coordinate.
The transverse distance between the \sigmamin\ beam track  and
the reconstructed vertex position was required to be less than $6 \sigma_3$,
where $\sigma_3 \approx 25 \rm \mu m$ typically. Furthermore events were 
rejected for which the beam track was connected to an outgoing track.

\section{Results}

\begin{figure}[h]
\begin{center}
  \mbox{\epsfxsize=13cm\epsffile{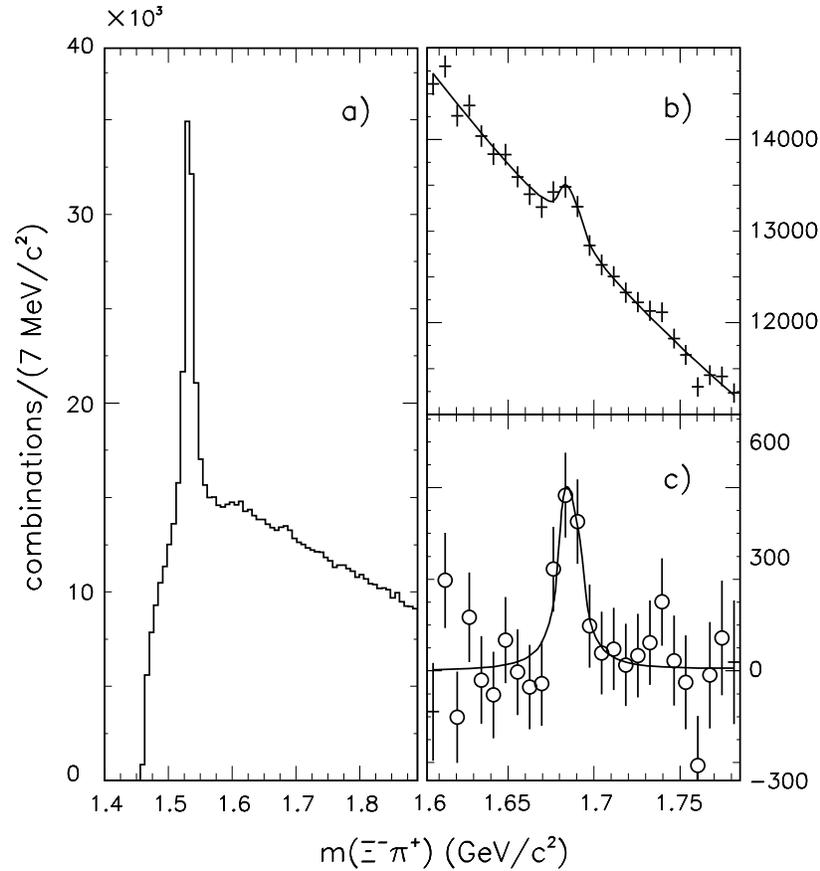}}
\vspace{0cm}
\caption{ Invariant mass distribution of the $\Xi^- \pi^+$ combinations.
          a)\ the \xipp\ and \xiss\ mass region; b) the \xiss\ mass region
            only; c) the \xiss\ mass region after background subtraction.}
\label{fig:mass1690}
\end{center}
\end{figure}

The $\Xi^- \pi^+$ mass distribution for all 
combinations of \ximin\ candidates with
positively charged particles  from the interaction vertex 
is presented in Figure 
\ref{fig:mass1690}a. This plot is dominated by the peak from \xipp\ decays.
The mass of the \xipp\  is measured to be $M = 1532.6  \pm 0.5$ \Mevc2\,
where the error comes mainly from the uncertainty
in the mass scale. It is in good agreement with the known value
$M = 1531.8 \pm 0.3$ \Mevc2\ \cite{pdg}.
The observed width is consistent with the known value of the
intrinsic width, $\Gamma = 9.1 \pm 0.5$ \Mevc2\ \cite{pdg}
and the mass resolution of our apparatus.
The number of \xipp\  decays is $63000 \pm 6000$,
where the error is mainly due to the uncertainties in the
shapes of the signal and background distributions.

Figure \ref{fig:mass1690}b
shows the mass region between 1600 and 1800 \Mevc2\ in more detail.
A resonance signal at about 1690 \Mevc2\ is visible above a large background.
For a more quantitative analysis the background was fitted with
a Legendre polynomial of second order in the mass range
from 1610 to 1792 \Mevc2 but excluding the resonance region.
In figure \ref{fig:mass1690}c the resonance signal is shown after
background subtraction.
Its mass and intrinsic width are
$$M = 1686 \pm 4 \Mevc2\ , \Gamma = 10 \pm 6 \Mevc2\ .$$
The number of observed events above background is
$1400 \pm 300$. Note that all quoted errors include uncertainties due
to reasonable variations of the signal and background shapes.
 
Finally we checked whether the observed signal could be
caused by a reflection from a $\Xi^- K^+$ state. 
We selected those combinations where the $\pi^+$ candidate
was actually positively identified as a $K^+$ by the RICH detector.
This sample contains about 4\% of the total sample and 
shows no resonant structure in the $\Xi^- \pi^+$ mass spectrum.

To measure the product of the production cross section and branching ratio,
$\sigma \cdot BR$, for \xiss\ relative to \xipp , 
we determined the apparatus acceptances from a Monte Carlo calculation. 
Within the observable kinematic range, $0.1< x_F<1$, 
the ratio of acceptances $r_A$ is very close to unity as expected from the 
similar decay kinematics: 
$ r_A = A(\Xi^0(1690) \rightarrow \Xi^{-} \pi^+) / A(\Xi^0(1530) 
   \rightarrow \Xi^{-} \pi^+)  = 0.98 \pm 0.02$. 
Within the large statistical uncertainties, the \xf\ distribution of the
\xiss\ is consistent with that of the \xipp .
We therefore assume them to be equal and obtain within the range of
$0.1< x_F<1$ a ratio for the $\sigma \cdot BR$ values of
\begin{equation}
\frac{\sigma \cdot BR (\Xi^0(1690) \rightarrow \Xi^{-} \pi^+) }
     {\sigma \cdot BR (\Xi^0(1530) \rightarrow \Xi^{-} \pi^+) } = 
0.022 \pm 0.005 .
\label{eq1}
\end{equation}

This number should be corrected for contributions from the
$\Xi^-$ admixture to the beam. 
The $\Xi^-$ flux was measured to be $(1.3 \pm 0.1)\%$ of the $\Sigma^-$ flux.
The production rate of the $\Xi^- (1320)$ by $\Xi^-$ is about equal to
the production rate by $\Sigma^-$ at $x_f \approx 0$ and enhanced by one
order of magnitude at $x_F \approx 0.5$ \cite{expt}.
The latter feature is related to the different numbers of common 
valence quarks in the projectiles and the produced $\Xi^- (1320)$. 
As expected from the smaller difference in the quark-overlap between a
$\Xi^-$ and a $\Xi^0(1530)$ on one hand and a $\Sigma^-$ and $\Xi^0(1530)$ 
on the other hand, a less pronounced enhancement is observed
for the $\Xi^0(1530)$ production \cite{Kel96}. 
We, therefore, expect that contaminations from $\Xi^-$ induced 
reactions to the observed $\sigma \cdot BR$ values are of the order 
of a few percent. Considering furthermore that contributions to 
the numerator and denominator in eq. \ref{eq1} partially cancel, 
we neglect them here.
 
\section{Discussion}

Our measured values for the mass and width of the \xiss\
are in reasonable agreement with the result of a
coupled channel analysis of $\Sigma^+K^-$ and 
$\Lambda\bar{K}^0$ spectra measured in K$^-$p interactions \cite{DIO78}.
That analysis provided first evidence for a $\Xi^0$ resonance with a mass
of 1684$\pm$5 \Mevc2 and a width of 20$\pm$4 \Mevc2. 
Both measurements of the $\Xi^0(1690)$ mass are slightly below the most 
significant value available for the
$\Xi^-(1690)$ mass, $m=1691.1\pm 1.9 \pm 2.0$ \Mevc2 \cite{BIA87I}.
Such a difference is in line with the mass splittings observed 
in the $\Xi(1320)$ and $\Xi(1530)$ systems of
6.4$\pm$0.6 and 3.2$\pm$0.6 \Mevc2 \cite{pdg}, respectively. 
We also note that the width of the 
$\Xi^0(1690)$ determined in the present experiment is comparable to
that of the \xipp .

From fig. 7 of ref. \cite{BIA87I}, an upper limit on the 
diffractive production cross section 
and branching ratio relative to \xipp\
can be estimated, $\sigma \cdot BR [\Xi^0(1690) \to 
\Xi^-\pi^+] / \sigma \cdot BR [\Xi^0(1530) \to \Xi^-\pi^+]  <0.03$.
The suppression factor of $0.022 \pm 0.005$ observed in the present
study is consistent with that upper limit.
For comparison it is interesting to note that in $\Xi^-$ induced interactions the ratio of
$\Xi^0 (1530)$ to $\Xi^0(1320)$ production was measured 
to be 0.25 at $x_F \approx 0.4$ and to increase with \xf\ \cite{schneider89}. 
It remains to be seen how much of the significantly stronger suppression found
in the present experiment is due to the opening up of the
$\Lambda \overline{K}$ and $\Sigma \overline{K}$ decay channels.

In conclusion we would like to point out that this is the first unambiguous
observation of the neutral member of the $\Xi (1690)$ doublet,
confirming the existence of this resonance at a mass of 1690 \Mevc2.

\section*{ Acknowledgements }

We are indebted to J.~Zimmer and the late Z.~Kenesei
for their help during all moments of detector construction and setup.
We are grateful to the staff of CERN's
EBS group for providing an excellent
hyperon beam channel, to the staff of CERN's Omega group for their
help in running
the $\Omega $-spectrometer and also to the staff of the SPS for providing
very good beam conditions.
We thank D.M.~Jansen, B.~Kopeliovich, O.~Piskounova and R.~Ransome for many
helpful discussions.

\end{document}